\documentclass[aps,prb,twocolumn,superscriptaddress]{revtex4-2}

\usepackage{amsmath}
\usepackage{amssymb}
\usepackage{sistyle}

\usepackage{hyperref}
\hypersetup{
	colorlinks=true,
	allcolors=blue,
	linktocpage=true,
	pdfhighlight=/N
}

\begin{document}

\title{Sagnac effect in a rotating ring with Dirac fermions}

\author{A.~Yu.~\surname{Fesh}}
\affiliation{Kyiv Academic University, 03142 Kyiv, Ukraine}

\author{Yu.~V.~\surname{Shtanov}}
\email{shtanov@bitp.kyiv.ua}
\affiliation{Bogolyubov Institute for Theoretical Physics, National Academy of Sciences of Ukraine, 03143 Kyiv, Ukraine}
\affiliation{Astronomical Observatory, Taras Shevchenko National University of Kyiv,  04053 Kyiv, Ukraine} %

\author{S.~G.~\surname{Sharapov}}
\email{sharapov@bitp.kyiv.ua}
\affiliation{Bogolyubov Institute for Theoretical Physics, National Academy of Sciences of Ukraine, 03143 Kyiv, Ukraine}
\affiliation{Kyiv Academic University, 03142 Kyiv, Ukraine}

\date{\today}

\begin{abstract}
The observation of the Sagnac effect for massive material particles offers a significant enhancement in sensitivity when compared to optical interferometers
with equal area and angular rotation velocity. For this reason, there have been suggestions to employ solid-state interferometers that rely on semiconductors
and graphene.
We investigate the Sagnac effect in Dirac materials governed by the relativisticlike quasiparticle dispersion law and show that the fringe shift is still determined by the mass of a free electron. This confirms that graphene is indeed a promising material for creating solid-state Sagnac interferometers. Considering monolayer graphene with its linear dispersion law and comparing it with light provides a deeper understanding of the Sagnac effect.
\end{abstract}

\maketitle

{\it Introduction.\/}
Physical phenomena associated with rotation possess a captivating allure
that spans multiple levels.
One of their fundamental illustrations  lies in the impossibility of establishing a standard clock synchronization procedure along a closed curve in a rotating material
(see Refs.~\cite{Landau.book2, Ryder2009book, Goy1997}).
One of the consequences of this impossibility is the renowned Sagnac effect, which
refers to the phenomenon where a phase shift
is observed between two coherent beams that travel along opposite paths
within an interferometer situated on a rotating disk
(see Refs.~\cite{Post1967RMP,  Stedman1997RPP,  Malykin2000UFN,  Pascoli2017} for reviews).
This phase shift, which was first demonstrated for light
by Sagnac in  \cite{Sagnac1913a, Sagnac1913b}, is an intrinsically relativistic effect.
It is not restricted to light waves,
but is observed for electron waves in vacuum \cite{Hasselbach1993PRA},
neutrons \cite{Werner1979PRL} and atoms \cite{Riehle1991PRL}
(see also the recent work in \cite{Gautier2022SciAd}).
Moreover, the observation of the Sagnac phase shift for massive particles
in solids, specifically superconducting Cooper pairs, dates back to as early
as the work in \cite{Zimmerman1965PRL}.

While practical applications of the Sagnac effect currently rely on light waves,
there is a compelling physical reason  why massive particles are
significantly more advantageous for its realization.
The Sagnac fringe shift, denoted by $\Theta_{\mathrm{S}}$,
with respect to the fringe position for the stationary interferometer
reads \cite{Malykin2000UFN, Hasselbach1993PRA, Hendricks1990QO, *Hendricks1990QO-corr}
\begin{equation}
\label{Sagnac-generic}
\Theta_{\mathrm{S}} = \frac{4 E A \Omega}{\hbar c^2} \, .
\end{equation}
This formula is applicable to waves comprising both massless and massive particles,
with $E$ being the total energy of a corresponding particle.
The quantity $A$ denotes the area of the enclosure, formed by the light or particle beams in the interferometer, orthogonally projected to the plane of rotation,
$\Omega$ is the angular velocity of rotation of the interferometer in the inertial frame, $\hbar$ is the reduced Planck constant, and $c$ is the
vacuum speed of light. This equation neglects
a small relativistic correction (discussed below).
Substituting in Eq.~(\ref{Sagnac-generic}) the energy
$E = \hbar \omega$, where $\omega$ is either the frequency of light
or the frequency of the de~Broglie wave of a material particle,
we recover the standard formula for the Sagnac phase shift  \cite{Post1967RMP, Stedman1997RPP, Malykin2000UFN, Hasselbach1993PRA, Pascoli2017}
\begin{equation}
\label{Sagnac-standard}
\Theta_{\mathrm{S}} = \frac{4 \omega A \Omega}{ c^2} \, .
\end{equation}

When considering light in vacuum and using the dispersion relation
$\omega = 2 \pi c /\lambda$, where $\lambda$ denotes the wavelength,
we arrive at another commonly used form
for the Sagnac phase shift
\cite{Post1967RMP,Stedman1997RPP,Malykin2000UFN},
\begin{equation}
\label{Sagnac-wavelength}
\Theta_{\mathrm{S}} = \frac{8 \pi  A \Omega}{c \lambda} \, .
\end{equation}

In the nonrelativistic case of
slow massive particles,
the energy $E = mc^2$ is associated with their rest mass $m$, and the phase fringe takes the form
\begin{equation}
\label{Sagnac-massive}
\Theta_{\mathrm{S}} = \frac{4 m A \Omega}{\hbar}\, .
\end{equation}
Comparing the phase shift for the matter-wave and optical interferometers,
we find that, for the same (projected) area and angular velocity,
the phase shift is enhanced by a factor  \cite{Clauser1988PC}
$m c^2 / \hbar \omega = \omega_0/\omega$, where $\omega_0 = m c^2/\hbar$
is the de~Broglie frequency mentioned above.
Since  the frequency $\omega_0$ is not a commonly used quantity, it is convenient to express Eq.~(\ref{Sagnac-massive}) for electrons similarly to Eq.~(\ref{Sagnac-wavelength}), with the photon wavelength $\lambda$ replaced by the Compton wavelength $\lambda_\mathrm{C} = 2 \pi \hbar / m_e c \approx \SI{0.0243}{\text{\AA}}$, where $m_e$ is the electron mass. This representation simplifies the comparison between interferometers using light and matter and having the same area and angular velocity.

For atoms, the matter-wave interferometer is significantly
more sensitive to rotation, with the enhancement factor relative to light reaching the value of $10^{10}$.
While the existing optical gyroscopes
necessitate the utilization of either several kilometers of optical fiber or a substantial area to reach the necessary sensitivity, its  enhancement for matter waves
has led to proposals to utilize cold-atom interferometers in the search
for smaller signals beyond the Earth's rotation \cite{Gautier2022SciAd}.

For free electrons, the enhancement factor reaches the value of $10^6$, and,
as mentioned earlier, the Sagnac effect was observed
using an electron interferometer  in vacuum \cite{Hasselbach1993PRA}.
The possibility of realizing the Sagnac effect in the solid state by employing 
a serial array of mesoscopic ring-shaped electron interferometers was
discussed in Refs.~\cite{Zivkovic2008PRB, Search2009PRA, Toland2010PLA, Search2011patent}
(see also Ref.~\cite{Bishara2008} for discussion of the Sagnac interferometry in carbon nanotubes).

As highlighted in \cite{Toland2010PLA,Search2011patent},
graphene emerges as a promising material for electron interferometry,
owing to its extraordinary electronic properties.
Recent experiments  on the Aharonov-Bohm oscillations
in ring-shaped gated bilayer graphene \cite{Ensslin2022NanoLett} and
magic-angle twisted bilayer graphene \cite{Iwakiri2023NatCom}
provide further confirmation of this potential
(see also the review in \cite{Chakraborti.review}).

However, monolayer graphene belongs to a new class of Dirac materials with zero effective carrier mass
and a linear dispersion relation similar to that of light. This unique behavior of charged carriers in graphene often prompts a reassessment of many fundamental principles of physics, leading to their deeper understanding. Thus, at first glance, due to the assertion discussed in Ref.~\cite{Malykin2000UFN}
that the phase shift remains independent of the wave's phase velocity, the Sagnac effect in graphene might appear analogous to the case of light. Another issue is that
the results discussed above pertain to massive particles in vacuum rather than inside solids.

The objective of the present work is to examine the Sagnac effect in graphene. More broadly,
the goal is to investigate the Sagnac effect in relativistic Dirac materials and to clarify its distinctions from the previously known cases.

{\it Model.\/}
We consider the Dirac materials characterized by the dispersion relation
\begin{equation}
\label{dispersion-law}
E({\boldsymbol k}) = \pm \sqrt{\hbar^2 v^2 k^2 + \Delta^2} +\mathcal{E}_\mathrm{D} \, ,
\end{equation}
where ${\boldsymbol k}$ represents the wave vector counted from the Dirac point,
which  can be either two- or three-dimensional, and the signs correspond to energy bands above and below this point. Below we will be examining a narrow ring of graphene on the plane geometry; in this case, the $z$ component of ${\boldsymbol k}$
can effectively be disregarded by setting it to zero. The quantity $v$ in Eq.~(\ref{dispersion-law})
denotes the Fermi velocity, and $\Delta$ is the gap in the quasiparticle spectrum.

The energy $\mathcal{E}_\mathrm{D} $ determines the energy position of the
Dirac point,  $E \left( {\boldsymbol k}=0 \right)$, with respect
to a chosen reference point.
For example, when considering the work function, the reference point is taken to be the local vacuum energy (see, e.g., Fig.~1 in \cite{Shubnyi2019PRB}). As we will see below, in the case of the Sagnac effect, it is crucial to include the electron rest energy $m_e c^2$ in $\mathcal{E}_\mathrm{D} $.

The gap term $\Delta$ is present, in particular, in the Hamiltonian
derived by Wolf (see Ref.~\cite{Fuseya2015JPSJ} for a review) for
Bi  and in similar effective Hamiltonians describing other three-dimensional 
(3D) Dirac materials.
It can also be induced in a graphene monolayer by placing it on top of hexagonal
boron nitride.

As already mentioned, the quasirelativistic spectrum (\ref{dispersion-law})
follows, for example, from the Wolf Hamiltonian as well as from
other effective low-energy 2D or 3D Dirac Hamiltonians used to describe
Dirac materials (for graphene, see, e.g., the review in \cite{Gusynin2007review}).

To focus on the Sagnac effect for quasiparticles with a
relativisticlike dispersion, we restrict ourselves to considering
the squared Dirac Hamiltonians neglecting the coupling between pseudospin and spin degrees
of freedom with the rotation of the frame
\cite{Mashhoon1988PRL}. Thus, we assume that a free quasiparticle in the
electron subsystem of the Dirac material in the inertial rest
frame $(t^\prime, {\boldsymbol r}^\prime )$ of the material
obeys the following wave equation:
\begin{equation}
\label{rest-eq}
\left( \hbar \frac{\partial}{\partial t'} + i \mathcal{E}_\mathrm{D} \right)^2 \psi -
v^2 \hbar^2 \nabla_{{\boldsymbol r}'}^2 \psi + \Delta^2 \psi = 0 \, .
\end{equation}
Here $\psi \left( t', {\boldsymbol r}' \right)$ is the electron envelop wave function,
which can be treated as a scalar.
Seeking a solution of Eq.~(\ref{rest-eq}) in the form
$\psi \sim \exp \left( - i E t'/\hbar +i {\boldsymbol k} {\boldsymbol r}' \right)$,
we reproduce the spectrum  (\ref{dispersion-law}).
After relating the energy gap $\Delta$ to the mass, $m = \Delta/v^2$,
and setting $v=c$ and $\mathcal{E}_\mathrm{D} =0$, Eq.~(\ref{rest-eq}) reduces to
the usual relativistic Klein-Gordon-Fock (KGF) equation.

{\it Sagnac effect from the KGF equation in the rotating frame.\/}
Before proceeding with the general analysis,
it is instructive, for an introspection,
to recapitulate a simple derivation of Eq.~(\ref{Sagnac-generic}) or (\ref{Sagnac-standard}) and
Eq.~(\ref{Sagnac-massive}) presented in Ref.~\cite{Hendricks1990QO}.
Essentially, it can be regarded as an extension of
the explanation of the Sagnac effect proposed
by Langevin in the framework of general relativity in 
\cite{Langevin1921}  (see also Refs.~\cite{Post1967RMP, Pascoli2017} for reviews  as well 
as \cite{Landau.book2, Ryder2009book}) to the case of relativistic wave equations.

The transformation from the primed rest frame to a nonprimed frame
rotating about the $z$ axis with angular velocity $\Omega$, in cylindrical coordinates, reads
\begin{equation}
\label{2rotating-frame}
t' = t\, , \quad r' = r \, , \quad \phi' = \phi +\Omega t \, .
\end{equation}
The KGF equation in the rotating frame was obtained in \cite{Hendricks1990QO} and it amounts to the replacement $\partial_{t'} \to \partial_t -
{\boldsymbol V} \cdot \nabla_{\boldsymbol r}$, where
${\boldsymbol V} = {\boldsymbol \Omega} \times {\boldsymbol r}$ is the local rotation velocity.

We consider wave signals, labeled by $\pm$, propagating in the counterclockwise
(in the angular direction $\phi$) and clockwise directions, respectively, around
a circle of radius $R$.
These signals are described by solutions of the KGF equation in the
rotating frame, $\psi_{\pm}(t,\phi) = \exp \left[ -i S_{\pm}(t,\phi) \right]$,
with phases
\begin{equation}
\label{sol-rotated-c}
S_{\pm}(t,\phi) = \omega t/\gamma - \gamma
\left(\Omega \omega R/c^2 \pm  k \right) R \phi \, ,
\end{equation}
where the Lorentz factor $\gamma = \left(1- \Omega^2 R^2 /c^2 \right)^{-1/2}$.
The frequency $\omega$ at the source in the rotating frame and
the wave number $k$ in Eq.~(\ref{sol-rotated-c}) are related by the
KGF dispersion relation $\omega^{2} = c^2 k^{2} + m^2 c^4/\hbar^2$.
Let us assume that the two counterpropagating signals originate in phase
from the same source at $\phi = 0$. Subsequently, these signals interfere
after completing a full round-trip, with the counterclockwise signal  at
$\phi = 2 \pi$ and the clockwise one at $\phi = -2 \pi$.
The phase difference between the two signals is therefore
\begin{equation}
\label{Sagnac-KGF}
\Theta_{\mathrm{S}} = S_{-}(t, - 2 \pi) - S_{+}(t, 2 \pi) =
\frac{4 \pi \gamma \omega \Omega  R^2}{c^2} .
\end{equation}
Considering that the circular interferometer has area $A = \pi R^2$,
we find that the last expression reproduces
Eq.~(\ref{Sagnac-standard})
up to the relativistic factor $\gamma$.
Nevertheless, this distinction is not crucial for our discussion, as $\Omega R \ll c$ by several orders
of magnitude, rendering the corrections to the fringe shift arising from $\gamma$ (which are of the order $\Omega^2 R^2/c^2$) practically indiscernible in experimental observations.

In the nonrelativistic limit, the phase fringe shift described
by Eq.~(\ref{Sagnac-massive}) emerges
as the KGF equation reduces to the Schr\"{o}dinger equation.
This can be done by transforming away the quickly oscillating
rest-energy dependence $\psi = \chi \exp \left( - i mc^2 t/\hbar \right)$, deriving the Schr\"{o}dinger equation for the slow factor $\chi$ in the rotating frame:
\begin{equation}
\label{KGF-2-Schrodinger}
i \hbar \frac{\partial \chi}{\partial t}=
\frac{1}{2m} \left(- i \hbar \nabla_{\boldsymbol r} - m  {\boldsymbol V}  \right)^2 \chi-
\frac{1}{2} m {\boldsymbol V}^2 \chi \, .
\end{equation}
Obtaining the solution of Eq.~(\ref{KGF-2-Schrodinger})
for the circularly propagating waves \cite{Hendricks1990QO}, one then finds the phases of the waves $\chi_\pm$ propagating in opposite directions:
\begin{equation}
\label{phase-Schrodiger}
S_{\pm}(t,\phi) = \left(\frac{\hbar k^{ 2}}{2m} - \frac{m \Omega^2 R^2}{2 \hbar} \right) t -
\left( \frac{m \Omega R}{\hbar} \pm k \right) R \phi \, .
\end{equation}
This equation can also be derived directly by substituting the nonrelativistic limit of the KGF dispersion relation, given by $\omega \approx mc^2/\hbar + \hbar k^{2}/ 2m$, into (\ref{sol-rotated-c}) and subtracting the relativistic contribution $mc^2/\hbar$. It is easy to see that Eq.~(\ref{phase-Schrodiger}) results in the phase fringe (\ref{Sagnac-massive})
(see also the discussion in Ref.~\cite{Vignale1995PL}).

The reason why the Sagnac effect survives in the nonrelativistic theory \eqref{KGF-2-Schrodinger} of the 
free particle lies in the fact that nonrelativistic quantum mechanics contains relativistic elements: 
Its wave function is not transformed as a scalar under Galilean transformations \cite{Hendricks1990QO, Dieks1990AJP}. Although the nonrelativistic phase \eqref{phase-Schrodiger} does not contain explicitly the de~Broglie frequency $\omega_0 = m c^2 / \hbar$ present in the phase of the relativistic wave function $\psi$, it is this frequency that enters the angular term of  (\ref{phase-Schrodiger}), responsible for the Galilean transformation of wave number
and generating the phase difference between the two counterpropagating waves.
In this regard, it is worth recallling that the original consideration of the matter waves by de~Broglie relied on the special relativity and included the de~Broglie frequency associated with particles at rest.

Before we return to investigating the Sagnac effect based on Eq.~(\ref{rest-eq}), the following points should be emphasized.
This equation for an envelope wave function arises from the Schr\"{o}dinger equation describing atomic orbitals, while the latter derives from a genuine Dirac equation describing electrons interacting with ions in solids.
For the spinless case, this step corresponds to the connection between the KGF equation and Eq.~(\ref{KGF-2-Schrodinger}).
There is no need for us to follow these steps in detail; it is sufficient to take into account that the wave functions of electrons in solids contain
the rapidly oscillating factor $\exp \left(- i mc^2 t/\hbar \right)$.
Although it is not observable under most circumstances and, therefore, is usually dropped, this is not the case for the Sagnac effect, for which the presence of this phase factor is crucial. Considering then Eq.~(\ref{rest-eq}), 
we should set $\mathcal{E}_\mathrm{D} =  mc^2$ to ensure the presence of this frequency in the solutions. This equality is valid up to the binding energy of electrons in solids, which is significantly smaller than their rest energy.

{\it Sagnac effect for the Dirac quasiparticles.\/}
With the above remarks in mind, we now turn to the Sagnac effect for the Dirac quasiparticles described by Eq.~(\ref{rest-eq}), where their velocity $v < c$.
It is clear that the approach of Ref.~\cite{Hendricks1990QO} described above for the KGF equation cannot be directly applied to the case of $v \neq c$. Indeed, we are dealing now with waves in a moving (rotating) medium, rather than in vacuum, and we cannot simply make the coordinate transformation \eqref{2rotating-frame} in Eq.~(\ref{rest-eq}), which describes waves in the material at rest.

Therefore, here we employ a different method, based on a covariant equation describing propagation of waves in a moving medium. To derive such an equation, we assume that the internal local structure (the crystalline lattice) of the material in motion is not sensitive to the arising small acceleration and that Eq.~\eqref{rest-eq} is satisfied in any small region of material in its instantaneous inertial rest frame. Since we treat the wave function $\psi$ as a scalar
[recall that we ignore the particle (pseudo)spin], it is easy to express all elements of Eq.~\eqref{rest-eq} in a covariant form in arbitrary space-time coordinates $x^\mu$. The partial derivative with respect to $t'$ (which is the proper time in the local inertial rest frame of an element of material) is covariantly expressed as $\partial_{t'} = u^\mu \nabla_\mu$, where $u^\mu$ is the element's four-velocity, and $\nabla_\mu$ is the covariant derivative (which, in particular, is just a partial derivative, $\nabla_\mu = \partial / \partial x^\mu$, when acting on a scalar). The spatial Laplacian in the local inertial rest frame of the element of material is most easily expressed as
\begin{equation}
\nabla_{{\boldsymbol r}'}^2 \psi = \partial^2_{t'} \psi - \Box \psi \, ,
\end{equation}
where $\Box = g^{\mu\nu} \nabla_\mu \nabla_\nu$ is the D'Alembertian.
Combining these expressions, we immediately write Eq.~\eqref{rest-eq} in arbitrary coordinates:
\begin{align} \label{covariant-eq}
& \left( u^\mu \nabla_\mu + i \mathcal{E}_\mathrm{D} \right)
\left( u^\nu \nabla_\nu + i \mathcal{E}_\mathrm{D} \right) \psi \nonumber \\
& - v^2 \left( u^\mu \nabla_\mu u^\nu \nabla_\nu - g^{\mu \nu} \nabla_\mu \nabla_\nu \right) \psi + \Delta^2 \psi = 0 \, .
\end{align}
Here and below we set the units $c = \hbar = 1$, unless stated explicitly otherwise. Since the small element of material is arbitrary, Eq.~\eqref{covariant-eq} is valid everywhere in the material in motion.
Obviously, for $v=1$ and $\mathcal{E}_\mathrm{D} =0$,  Eq.~(\ref{covariant-eq}) reduces to the KGF equation written in the covariant form. For material at rest, in its inertial rest frame, we have $u^0 = 1$, with the spatial components of $u^\mu$ being zero, and Eq.~\eqref{covariant-eq} reduces to Eq.~\eqref{rest-eq}, as required.

Similarly to the consideration of the KGF equation,
we look for a solution of Eq.~(\ref{covariant-eq}) in the form $\psi = {\cal A} e^{- i S}$, where both the amplitude ${\cal A}$ and the phase $S$ can depend on coordinates. Employing the eikonal approximation \cite{Landau.book2}, i.e., neglecting the second derivatives of the amplitude, from \eqref{covariant-eq}
we obtain the following equation for the phase:
\begin{align}\label{Jacobi}
- v^2 g^{\mu \nu} \nabla_\mu S \nabla_\nu S - \left( 1 - v^2 \right) \left( u^\mu \nabla_\mu S \right)^2 \nonumber \\
{} + 2 \mathcal{E}_\mathrm{D} u^\mu \nabla_\mu S + \Delta^2 -
\mathcal{E}_\mathrm{D}^2 = 0 \, .
\end{align}
The explicitly covariant form of this equation allows us to
consider the problem in arbitrary space-time coordinates,
for instance, in the frame rotating with the sample, or in the nonrotating frame.

As in the case of the KGF equation, we consider a narrow ring of radius $R$ rotating with angular velocity $\Omega$.  In the non-rotating frame $\left( t', \phi' \right)$ (the radial coordinate is effectively frozen at $r' = R$ ),
the invariant interval and the contravariant metric tensor are
\begin{equation}
\label{interval-rest}
d s^2 =  d t^{\prime\, 2} - R^2 d \phi^{\prime\, 2}, \quad
g^{\mu\nu} =  \begin{pmatrix}
                  1 &  0 \\
                  0 &  - R^{-2}
              \end{pmatrix} \, .
\end{equation}
Accordingly, in the rotating coordinate frame $(t, \phi)$ obtained by the coordinate transformation \eqref{2rotating-frame}, the invariant interval is
\begin{equation}
d s^{2} =  \gamma^{-2} d t^{2} - 2  \Omega R^2 d t d \phi  - R^{2} d \phi^{2} \, ,
\end{equation}
and the covariant and contravariant metric tensors read
\begin{equation}
\label{metric-rot}
g_{\mu\nu} = \begin{pmatrix}
					\gamma^{-2} &  -\Omega R^2\\
                  -\Omega R^2 &  -  R^{2}
			\end{pmatrix} \, ,
\quad
g^{\mu\nu} = \begin{pmatrix}
				1 &  -\Omega \\
                -\Omega &  - \gamma^{-2} R^{-2}
			\end{pmatrix} \, .
\end{equation}

To draw parallels to the case of the KGF equation, we begin by examining Eq.~(\ref{Jacobi})
in the rotating frame. Since the ring is at rest in this
frame, the components of the four-velocity $u^\mu$ in coordinates $(t, \phi)$ are
$u^\mu = \left( \gamma, 0 \right)$, obeying the usual normalization condition $g_{\mu\nu} u^\mu u^\nu =1$.
Again, let $S_\pm$ be the phases of waves propagating along ($+$) and opposite ($-$)
to the angular direction $\phi$. For a positive-frequency solution, we then have
\begin{equation}\label{phase-rot}
S_\pm = \omega_\pm t - k_\pm R \phi \, ,
\end{equation}
where $\omega_\pm$ and $k_\pm$ are the corresponding constant frequencies and wavenumbers.
The source and receiver of electrons move together with the ring with the same four-velocity $u^\mu = d x^\mu (\tau) / d \tau$, where $\tau$ is the proper time, by assumption, detecting the same frequency $\omega$ for both waves:
\begin{equation}
\label{frequency-fix}
\omega = \frac{d}{d \tau}S_\pm (x(\tau)) = u^\mu \nabla_\mu S_\pm = \gamma \omega_{\pm} \, .
\end{equation}

Substituting \eqref{phase-rot} and \eqref{frequency-fix} into \eqref{Jacobi} and using the relations $\nabla_t S_\pm = \omega_{\pm}$ and $\nabla_\phi S_\pm = - k_{\pm} R$,
after some simple calculations, we obtain the following equation:
\begin{equation}
\label{k-vs-omega}
v^2 \left( \frac{k_\pm}{\gamma} - \omega \Omega R \right)^2 -
\left( \omega - \mathcal{E}_\mathrm{D} \right)^2 + \Delta^2 = 0 \, .
\end{equation}
Solving this equation with respect to $k_{\pm}$,
we find
\begin{equation}\label{kpm}
k_\pm = \gamma \omega \Omega R  \pm \frac{\gamma}{v}
\sqrt{\left( \omega - \mathcal{E}_\mathrm{D}  \right)^2  - \Delta^2} \, .
\end{equation}
Note that the main contribution to the wavenumber contains the velocity constant $v$, but its correction induced by rotation does not depend on this velocity.
Furthermore, one can easily see from the dispersion law (\ref{dispersion-law}) with $\omega = E({\boldsymbol k})/\hbar$
that the second term on the right-hand side of Eq.~(\ref{kpm}) is nothing but the wave vector $\gamma k$.
Thus, the solution (\ref{phase-rot}) with $\omega_{\pm}$ and $k_{\pm}$ given by
Eqs.~(\ref{frequency-fix}) and (\ref{kpm}), after restoring $c$ and $\hbar$,
formally coincides with the previously presented solution of the KGF equation (\ref{sol-rotated-c}).
The important difference is that the supporting wave obeys the dispersion law (\ref{dispersion-law}).
Accordingly, for the case of Dirac fermions under consideration, the Sagnac phase fringe is still determined by Eq.~(\ref{Sagnac-KGF}), but with the frequency $\omega = E({\boldsymbol k})/\hbar$ characterized by the
dispersion law (\ref{dispersion-law}).

It is clear that the energy $\mathcal{E}_\mathrm{D}$ in Eq.~(\ref{dispersion-law})
is at least $10^5$ times larger than the dispersive part. Thus we conclude that
the Sagnac effect in the Dirac materials either with $\Delta=0$ or with finite $\Delta$
is characterized by Eq.~(\ref{Sagnac-massive}) with $m \approx m_e$,
that is, the wave supporting the Sagnac effect has
the same nature as the waves in existing electron interferometers
\cite{Ensslin2022NanoLett,Iwakiri2023NatCom}. The corresponding wave numbers
are equal to the Fermi wave vector $k_\text{F}$. We emphasize that it is incorrect to use Eq.~(\ref{Sagnac-wavelength}) with the Fermi wavelength $\lambda_\text{F} = 2\pi/k_\text{F}$
for the Sagnac effect in this case. It is the de~Broglie frequency $\omega_0$ and the
corresponding Compton wavelength $\lambda_\text{C}$ that determine the phase fringe.

To illustrate the covariant character of our approach,
we reproduce the results obtained above in the non-rotating coordinate frame
$(t^\prime, \phi^\prime)$. The four-velocity in this frame acquires the spatial (angular) component
$u^\mu = \left( \gamma , \gamma \Omega \right)$, satisfying the normalization condition $g_{\mu\nu} u^\mu u^\nu =1$ with the metric given by Eq.~(\ref{interval-rest}).
We can look for the phases $S_{\pm}$ of the waves moving in opposite directions in the same form as in Eq.~(\ref{phase-rot}),
\begin{equation}\label{phase-rest}
S_\pm = \omega'_\pm t' - k'_\pm R \phi' \, ,
\end{equation}
where $\omega'_\pm$ and $k'_\pm$ are the corresponding constant frequencies
and wave numbers in the nonrotating frame. The condition $u^\mu \nabla_\mu S_\pm = \omega$
[see Eq.~(\ref{frequency-fix})] in this frame reads
\begin{equation}\label{frequency-fix-rest}
\gamma \left( \omega'_\pm - k'_\pm \Omega R \right) = \omega \, .
\end{equation}
Equation~(\ref{Jacobi}) for $S_{\pm}$, using Eq.~(\ref{frequency-fix-rest}) and the condition $u^\mu \nabla_\mu S_\pm = \omega$, takes the form
\begin{equation}\label{k-vs-omega-pm-rest}
v^2 \left( k^{\prime\, 2}_\pm - \omega^{\prime\, 2}_{\pm} \right)  -
\left( 1 - v^2 \right) \omega^2 + 2 \mathcal{E}_\mathrm{D}  \omega
+ \Delta^2 - \mathcal{E}_\mathrm{D}^2 = 0 \, .
\end{equation}
Equation~(\ref{frequency-fix-rest}) shows that the frequencies $\omega'_\pm$ are Doppler
shifted relative to each other in the nonrotating frame. Expressing $\omega'_\pm$ via $k'_\pm$ from \eqref{frequency-fix-rest} and substituting them into Eq.~(\ref{k-vs-omega-pm-rest}), we obtain
an equation identical to Eq.~(\ref{k-vs-omega}), but with $k'_\pm$ instead of
$k_\pm$. This implies that $k'_\pm = k_\pm$, with the wavenumbers previously determined in Eq.~(\ref{kpm}).
Thus, we have found the phases $S_{\pm}$ in Eq.~(\ref{phase-rest}).
Finally, we can can proceed to the rotating frame in the solution by the coordinate transformation
(\ref{2rotating-frame}). In this frame, the phases (\ref{phase-rest}), using the expressions for $\omega'_\pm$ found from \eqref{frequency-fix-rest}, become
$S_\pm = \omega t/\gamma - k_\pm R \phi$,
in full agreement with the previous result, Eqs.~(\ref{phase-rot}) and \eqref{frequency-fix}.

{\it Conclusion.}
To summarize, we have shown that the phase fringe for
the Sagnac effect in graphene and other Dirac materials, with the charge carrier dispersion
law given by Eq.~(\ref{dispersion-law}), is described by Eq.~(\ref{Sagnac-massive})
with $m$ being the free-electron mass (up to its binding energy in graphene). This is the same expression that was previously obtained for the Sagnac phase fringe
for electrons in vacuum. The existing experiments on the Aharonov-Bohm oscillations
in rings \cite{Ensslin2022NanoLett, Iwakiri2023NatCom} suggest the possibility of observing
the Sagnac effect in nonsuperconducting solid-state systems.

We would like to thank Iu.\,A.~Chernii, E.\,V.~Gorbar, V.\,P.~Gusynin, A.\,A.~Kordyuk, E.\,G.~Len, V.\,M.~Loktev,  T.-H.~Pokalchuk, A.\,O.~Slobodeniuk,
and A.\,A.~Varlamov for numerous stimulating and enlightening discussions.
We are grateful to the Armed Forces of Ukraine for providing security to perform this work.
A.Y.F. acknowledges support from the National Research Foundation of Ukraine (project 2020.02/0051).
Y.V.S. and S.G.S. acknowledge support from the
National Academy of Sciences of Ukraine 
under Projects 0121U109612 and 0122U000887, respectively,
and from the Simons Foundation (U.S.).

\bibliography{sagnac.bib}

\end{document}